\documentclass{aa}  
\usepackage{graphicx}
\usepackage{txfonts}
\usepackage{lscape}
\usepackage{graphicx}
\usepackage{epstopdf}
\usepackage{natbib}
\bibpunct{(}{)}{;}{a}{}{,}
\def\r14{$R^{1/4}$}

\def\kmsMpc{km\,s$^{-1}$\,Mpc$^{-1}$}

\def\mueff{\ifmmode{\mu_{\rm e}}\else{$\mu_{\rm e}$}\fi}
\def\zf{\ifmmode{z_{\rm f}}\else{$z_{\rm f}$}\fi}
\def\Ks{\ifmmode{K_{\rm s}}\else{$K_{\rm s}$}\fi}

\renewcommand{\deg}{\ensuremath{{^\circ}}}

\newcommand{\MR}{\ensuremath{{M}_\mathrm{R}}}
\newcommand{\MB}{\ensuremath{{M}_\mathrm{B}}}

\newcommand{\SExtractor}{\textsc{SExtractor}}
\newcommand{\Hyperz}{\textsc{HyperZ}}
\newcommand{\GimTwoD}{\textsc{Gim2D}}

\newcommand{\zspec}{\ensuremath{{z_\mathrm{spec}}}}

\newcommand{\zmax}{\ensuremath{{z_\mathrm{max}}}}

\newcommand{\RdeV}{\ensuremath{R^{1/4}}}

\begin{document}
   \title{Bulges of disk galaxies at intermediate redshifts.II.}

   \subtitle{Nuclear, disk and global colours in the Groth Strip}

   \author{Lilian Dom\'\i nguez-Palmero
          \and
          Marc Balcells}

   \offprints{L. Dom\'\i nguez-Palmero}

   \institute{Instituto de Astrof\'\i sica de Canarias, 
              E-38200 La Laguna, Tenerife, Spain\\
              \email{ldp@iac.es, balcells@iac.es}}

\date{\large \textit{Draft \today}}


 
\abstract{The chronology of bulge and disk formation is a major unsolved issue in galaxy formation, which impacts on our global understanding of the Hubble sequence.}
{We analyse colours of the nuclear regions of intermediate redshift disk galaxies, with the aim of obtaining empirical information of relative ages of bulges and disks at $0.1 < z < 1.3$.}
{We work with an apparent-diameter limited parent sample of 248 galaxies from the HST Groth Strip Survey. We apply a conservative criterion to identify bulges and potential precursors of present-day bulges based on nuclear surface brightness excess above the exponential profile of the outer parts and select a sample of 56 galaxies with measurable bulges. We measure bulge colours on wedge profiles opening on the semi-minor axis least affected by dust in the disk, and compare them to disk, and global galaxy colours.}  
{For $60\%$ of galaxies with bulges, the rest-frame nuclear colour distribution shows a red sequence that is well fit  by passive evolution models of various ages, while the remainder $40\%$ scatters towards bluer colours. In contrast, galaxies without central brightness excess show typical colours of star forming population and lack a red sequence. We also see that, as in the local Universe, most of the minor axis colour profiles are negative (bluer outward), and fairly gentle, indicating that nuclear colours are not distinctly different from disk colours. This is corroborated when comparing nuclear, global and disk colours: these show strong correlations, for any value of the central brightness prominence of the bulge. No major differences are found between the low and high inclination samples, both for the bulge and non-bulge samples.}  %
{Comparison with synthetic models of red sequence bulge colours suggests that such red bulges have stopped forming stars at an epoch earlier than $\sim 1$ Gyr before the observation. The correlation between nuclear and disk colours and the small colour gradients hints at an intertwined star formation history for bulges and disks: probably, most of our red bulges formed in a process in which truncation of star formation in the bulge did not destroy the disk.}



   \authorrunning{Dom\'\i nguez-Palmero \& Balcells}
   \titlerunning{Bulges of disk galaxies at intermediate redshifts. II.}

   \maketitle
%

\section{Introduction}
\label{sec:introduction}

Three reasons make bulge population ages useful for galaxy formation studies.  In all inside-out scenarios, the bulge regions should contain information on the earliest phases of star formation in galaxies.  Also, when compared to ages of elliptical galaxies, bulge ages should shed light on whether bulges and ellipticals share a formation history as proposed in CDM-based semi-analytical models \citep{Baugh96,Kauffmann96bul}.  Finally, bulge population ages, when compared to ages of the disks they live in, should provide clues on whether bulges formed before disks, again as assumed in current CDM-based models, or after disks, as proposed in secular evolution models \citep{Kormendy04}.  

The current consensus emerging from local Universe studies points to bulges being uniformly old ($\sim$10 Gyr) in early- to intermediate-type galaxies.  \citet{Peletier99} show that optical-NIR colours are indistinguishable from those of ellipticals in Coma, suggesting similar ages of at most 2 Gyr younger than Coma ellipticals.  Independently, from stellar colour-magnitude diagrams rather than integrated colours, for the Milky Way bulge \citet{Zoccali03} infer homogeneous stellar ages similar to those of globular clusters, or $\ga$10 Gyr.  Bulges of later-type spirals have signs of harboring younger populations, and ongoing star formation; \citep[e.g.][]{Peletier99,Carollo01}.  

Studies of local galaxies have also shown that redder bulges live in redder disks \citep{Balcells94}, i.e., colour-wise, bulges are more similar to their parent disks than to each other \citep{Peletier96}. This suggests common evolution, although not necessarily that one forms from the other. 

Because the bulge ages just mentioned do not appear to be much older than the oldest stars in nearby mature disks  (about 10~Gyr for the MW disk), the relative chronology of disk and bulge formation appears difficult to determine from the fossil record in nearby galaxies.  Studying galaxies at cosmologically-significant look-back times offers a means to date galaxian components seen at epochs closer to that of their formation, and may provide better time resolution to determine age differences between bulges and ellipticals, and between bulges and disks. This approach has been followed by several groups in recent years, with conflicting conclusions. \citet{Ellis01} and \citet{Menanteau01} find a prevalence ($30 - 50\%$)  
of young field spheroids at $z \sim 1$ in the two HDFs.  In contrast, \citet{Koo05} emphasized that $85\%$ of luminous ($M_B < -19$) field bulges at redshifts $z \sim 0.8$ in the Groth strip are nearly as red ($U - B \sim 0.50$) as local E/SO's, a result that suggests both an old metal-rich dominant population and a 'drizzling' contribution of star formation to the global colours.  \citet{Koo05HDF} compared bulge colours for field and cluster galaxies, concluding that bulge colours are identical for field and cluster galaxies.   
\citet{MacArthur07}, who isolated bulges for a sample of 137 spiral galaxies within the redshift range $0.1<z<1.2$ in the GOODS fields, find that bulge colour is related to mass. The most massive bulges are as old and red as massive spheroids, which would be consistent with a mass assembly at high redshift. Smaller bulges, in contrast, have quite diverse star formation histories, with significant star formation at $z < 1$.  Additional, indirect evidence for late formation of galaxy bulges was presented by \citet{Hammer01}, who argued that some luminous, blue, compact sources at $0.5<z<1$ may be starburst phases leading to the formation of bulges of massive spirals.  

While cosmic variance likely affects the diversity of results presented above, most likely their differences are strongly affected by the way bulges are selected, the way bulge colours are measured, and the way bulge age information is inferred from the measured colours by the various studies. \citet{Ellis01} and \citet{Menanteau01} work with 'field spheroidals' selected on the basis of concentration and asymmetry indices, as well as from visual classification, while \citet{Koo05} define their samples of 'photobulges' on the basis of modeling surface brightness distributions as the sum of \r14\ and exponential components, using the \GimTwoD\ software \citep{Simard02}.  Both methods allow for pure ellipticals to make it into their samples, which may be seen as a draw-back.  Moreover, \citet{Ellis01} and \cite{Menanteau01} may include objects with purely exponential surface brightness profiles and exclude bulges from star-forming spiral galaxies. 
 
Colour measurements are carried out on each image pixel by \citeauthor{Menanteau01}, while a global photo-bulge colour is derived by \citeauthor{Koo05} from constrained bulge-disk fits on two bands.  Finally, \citet{Menanteau01} infer ages from the internal dispersion of galaxy colours, while \citet{Koo05} compare integrated rest-frame colours, and colour-magnitude diagrams, to those from local E-S0's and distant clusters.  

The above studies are, probably, complementary in the estimation of population ages of the precursors of galaxy bulges. In this paper we analyse colours of a sample of bulge galaxies, which has been defined in a companion paper \citep[hereafter Paper~I]{Dominguez08I}, with the aim of addressing similar questions to the mentioned studies. We aim to provide complementary clues to the formation of galaxy bulges, given our different approach.  Our methods are detailed in Paper~I.  Specifically, we work with a diameter-limited sample, and we define a prominence index $\eta$ to identify a subsample of galaxies with measurable bulges. Section~\ref{sec:data} sumarizes the approach we use to define the bulge sample and the method to measure colours. In Sect.~\ref{sec:comparison} we compare our bulge sample with that of \citet{Koo05}. In the remainder of the sections we discuss the redshift distribution of the bulge sample; the shape of the minor-axis colour profiles; rest-frame colours for bulges and non-bulge galaxies; and the comparison of bulge colours to global and disk colours.  

A cosmology with  $\Omega_M = 0.3$, $\Omega_\Lambda = 0.7$, $H_0 = 70$ \kmsMpc\ is assumed throughout.  Magnitudes are expressed in the Vega system.

\section{Data}
\label{sec:data}

Details of the approach adopted to define the bulge sample and to infer bulge, disk and global colours are presented in Paper I, as well as the tabulated characteristics (Tables A.1 and A.2), measured colours (Tables A.3 and A.4) and colour gradients (Tables A.5 and A.6) for the bulge galaxy sample and the complementary galaxy sample with no measurable bulges. Images, colour maps, photometric spectral energy distributions (SED), semi-major surface brightness profiles and semi-minor colour profiles of sample sources are also presented in Paper I, Appendix B. Colour profiles from the bluer  semi-minor axes, geometrically deprojected to face-on and scaled to kiloparsecs are also shown in Paper I, Appendix C, to facilitate comparison between profiles of bulge and non-bulge galaxies.

Briefly, we work with a 2.8\arcsec-diameter limited sample of galaxies from the HST Groth Strip Survey, divided in two subsamples with different inclinations: high inclination sample ($50\deg < i < 70\deg$) and low inclination sample ($i < 50\deg$), both spaning redshift $0.1 < z < 1.3$, spectroscopic and photometric. This division is made to assess the effects of dust in the measured quantities, as well as to have at least one subsample free from pollution by ellipticals (only E7 ellipticals are expected in the high-inclination subsample). From these initial samples we select bulge samples using a central light excess criterion: we define a \textsl{bulge prominence index} $\eta$ that measures the excess central surface brightness over the inward extrapolation of the exponential profile of the outer disk. We then consider that galaxies with measurable bulges are those with $\eta > 1$. The $\eta$ index is a sort of concentration index, but is more sensitive to small bulges in extended disks than standard concentration indices. Using this method, we are selecting not only bona fide bulges of early type spirals, but also possible progenitors of present-day bulges. We avoid also the bias caused by selection methods based on bulge-disk decompositions. These fit theoretical axi-symmetric bulge and disk models to galaxy images which often barely resolve the bulge; the irregular morphologies of many disks further contribute to making the fits unstable.  

To derive bulge colours, we extract colour profiles along wedge-shaped apertures opening on the minor axes. We chose this method to reduce reddening from dust in the disk, which can be quite important in inclined galaxies.  We avoid dust reddening by working on the bluest of the two semi-minor axis profiles. 
Other processes may yield asymmetric colour profiles on the two minor axes, such as an asymmetric distribution of nuclear star forming regions.  Also, if the bulge were much redder than the disk, and all the galaxy dust were placed in a thin, opaque mid-plane layer \citep[see][]{Tuffs04}, then it would be the redder, rather than the bluer side, which best matched the bulge colours.  
We argue, on the basis of the geometry of the dust lanes in the galaxies, that the latter configuration does not apply to the vast majority of our galaxies, and, as a result, the bluer of the two minor axes is the one that best approximates the intrinsic colours of the bulge populations.
We work with direct colour measurements over the colour profiles at a fixed metric distance (0.85 kpc) from the center; we avoid colour measurements derived from a bulge-disk decomposition of the galaxies images whose dependence on the bulge-disk modeling is difficult to predict. 

To extract disk colours, we average in the range from 1 to 2 scale lengths derived from the exponential law fit to the outer regions of the galaxies on the semi-major axis wedge-shaped aperture profile. Global galaxy colours are obtained using a fixed 2.6\arcsec\ circular aperture, on images smoothed to a common FWHM of 1.3\arcsec.
The K-corrections and colour transformations are calculated with SEDs obtained from the best-fit solution, delivered by \Hyperz\ \citep{hyperz}, to photometric points in the bands $U, B, F606W, F814W, J, \Ks$.

\begin{table*}
\caption{Median observed $(F606W-F814W)$ colour gradients and dispersions} \label{tab:colgradient}
\begin{center}
\begin{tabular}{l c c c c c c c c}
\hline
\hline
  & \multicolumn{4}{c}{$i < 50$} & \multicolumn{4}{c}{$50 < i < 70$}\\
\hline
 & \multicolumn{2}{c}{$\eta > 1$} & \multicolumn{2}{c}{$\eta < 1$} & \multicolumn{2}{c}{$\eta > 1$} & \multicolumn{2}{c}{$\eta < 1$}  \\
\hline
 & med & $\sigma$ & med & $\sigma$ & med & $\sigma$ & med & $\sigma$ \\
\hline
bluer smb & -0.014 & 0.032 & -0.027 & 0.073 & -0.032 & 0.036 & -0.025 & 0.051\\
redder smb & -0.019 & 0.023 & -0.026 & 0.081 & -0.021 & 0.045 & -0.022 & 0.055 \\
sma & -0.014 & 0.033 & -0.028 & 0.063 & -0.020 & 0.033 & -0.025 & 0.063 \\
\hline
\hline
\end{tabular}
\end{center}
\begin{footnotetext}
TNote.- Median observed $(F606W-F814W)$ colour gradients and standard deviation, in units of $\Delta(V-I)$ per kpc, calculated over the bluer semi-minor axis (bluer smb), over the redder semi-minor axis (redder smb) and over the averaged semi-major axis (sma) colour profiles, for samples with bulges ($\eta > 1$) and without bulges ($\eta < 1$) and for the low- ($i < 50$\deg) and high- ($50\deg < i < 70$\deg) inclination samples.
\end{footnotetext}
\end{table*}

\begin{figure}
\begin{center}
\includegraphics[angle=0,width=0.45\textwidth]{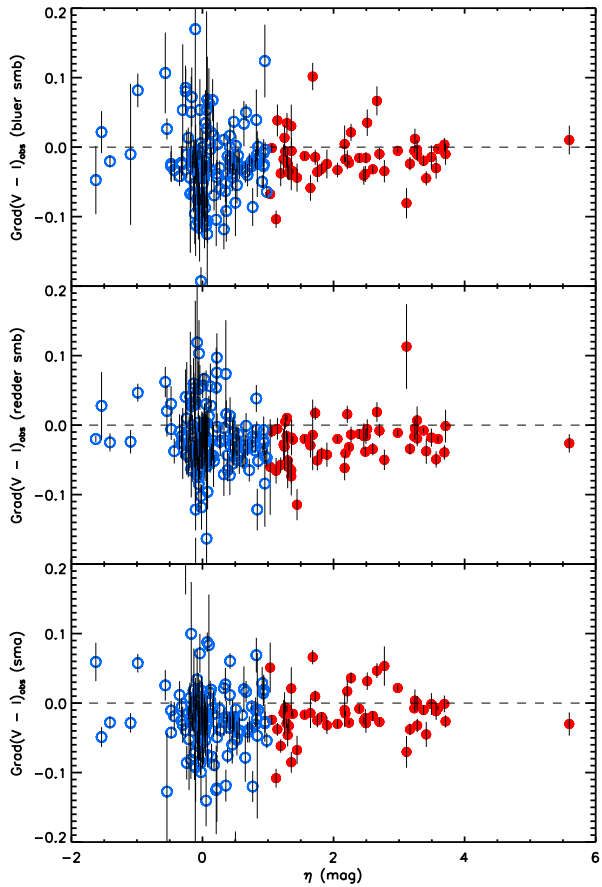}%
\end{center}
\caption{Colour gradients vs the central brightness excess, $\eta$. (a) Colour gradient calculated over the bluer semi-minor axis profile, (b) colour gradient calculated over the redder semi-minor axis profile, (c) colour gradient calculated over the averaged semi-major axis profile. {\it Filled circles} (red in the electronic edition): galaxies with prominent bulge ($\eta > 1$). {\it Open circles} (blue in the electronic edition): galaxies without bulges ($\eta < 1$).}   \label{fig:grad_eta}
\end{figure}

\begin{figure*}
\begin{center}
\includegraphics[angle=0,width=0.9\textwidth]{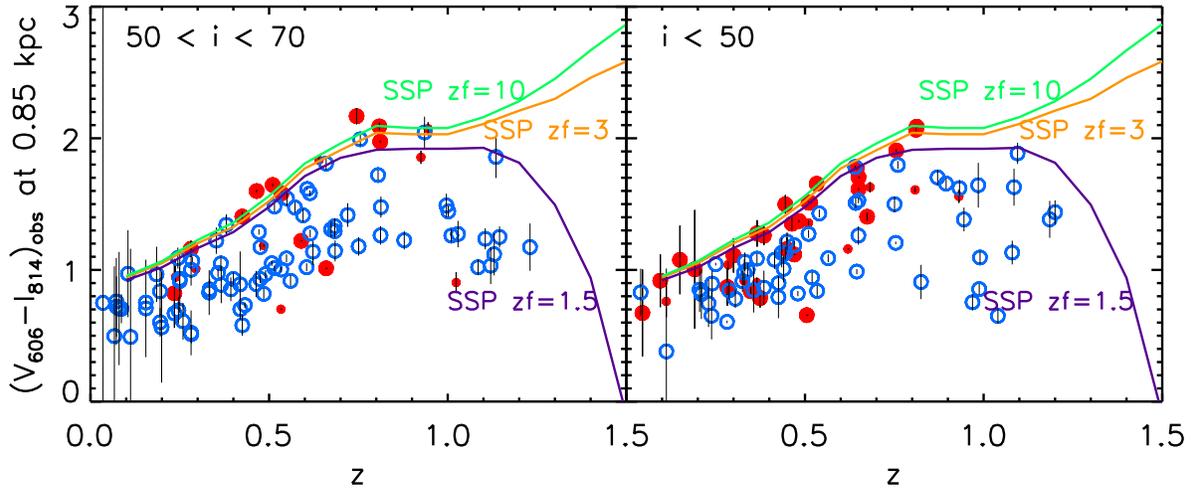}%
\end{center}
\caption{Observed $(F606W - F814W)$ colours measured at 0.85 kpc from the center along the minor axis vs redshift for the high- and low-inclination samples. {\it Filled circles} (red in the electronic edition): colours of galaxies with prominent bulge ($\eta > 1$). {\it Open circles} (blue in the electronic edition): colours of galaxies without bulges ($\eta < 1$). {\it Solid lines}: observed colour tracks for passive evolution systems (SSP model with solar metallicity, a \citet{Chabrier03} IMF, no dust) with formation redshifts equal to 10, 3 and 1.5, top to bottom.}   \label{fig:viobs085_z}
\end{figure*}

\begin{figure*}
\begin{center}
\includegraphics[angle=0,width=0.9\textwidth]{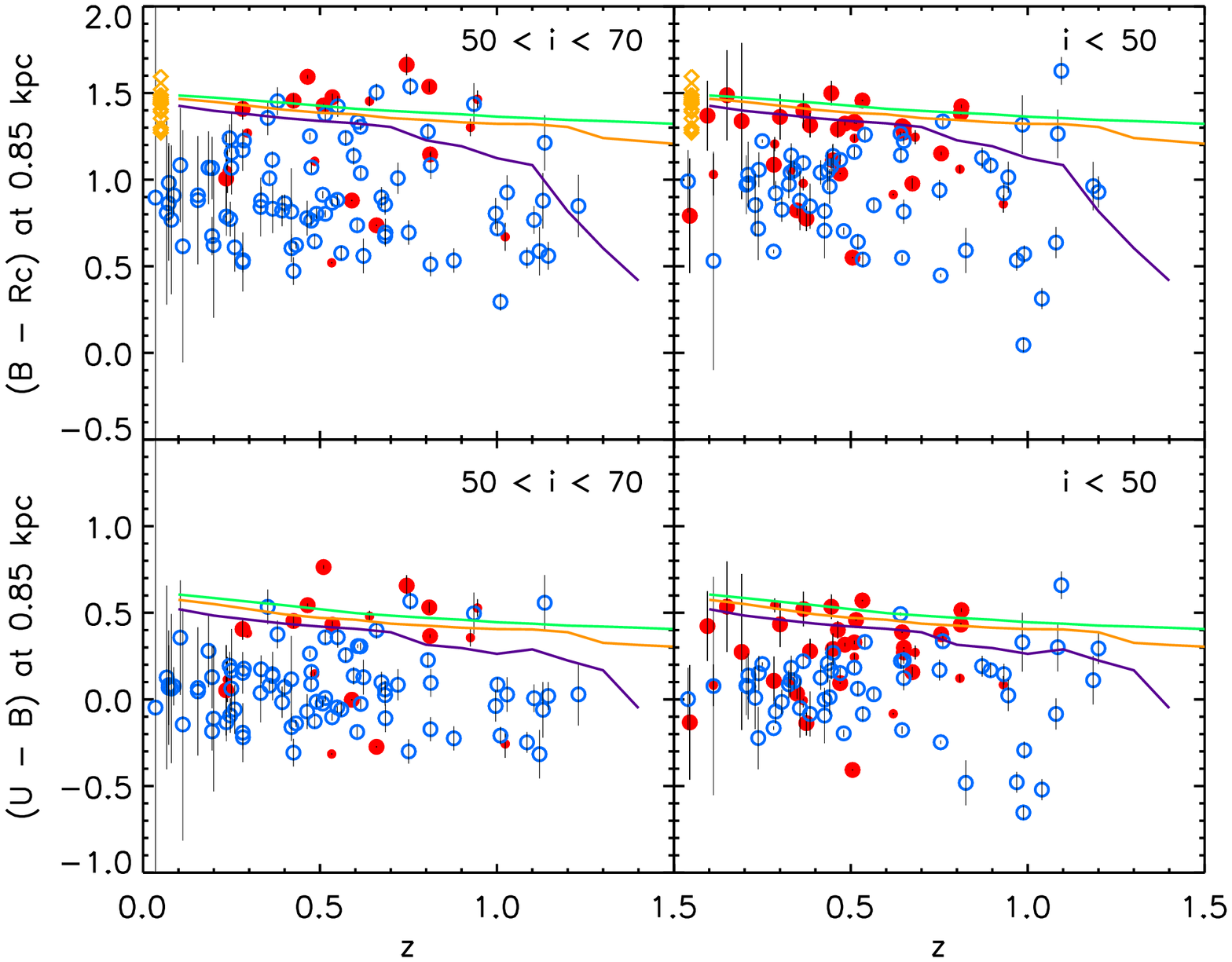}%
\end{center}
\caption{Rest-frame central $(B - R)$ and $(U-B)$ colours vs redshift for the high- (left panels) and low-inclination (right panels) samples. {\it Filled circles} (red in the electronic edition): colours of galaxies with prominent bulge ($\eta > 1$). {\it Smaller filled circles}: galaxies with prominence indices in the range $1 < \eta < 1.5$. {\it Open circles} (blue in the electronic edition): colours of galaxies without bulges ($\eta < 1$). {\it Orange diamonds} in upper pannels: rest-frame $(B-R)$ colours of local bulges \citep{Peletier96}. {\it Solid lines}: rest-frame colour tracks for passive evolution systems (SSP model with solar metallicity, a \citet{Chabrier03} IMF, no dust) with formation redshifts equal to 10, 3 and 1.5, top to bottom.}   \label{fig:br_z}
\end{figure*}


\subsection{Comparison sample}
\label{sec:comparison}

The analysis of bulge colours in the Groth strip by \citet{Koo05} provides an ideal benchmark for comparing not only results but also aspects of our methodology ranging from sample selection to colour determinations to K-corrections. 
\citet{Koo05}'s sample comprises 86 candidate bulges with redshifts from 0.73 to 1.04.  The sample was selected from the same HST GSS images we use, but the analysis follows different principles.  Selection is done purely on bulge magnitude ($F814W < 23.57$ in Vega system), which is derived from decomposition of the galaxy images into outer exponential ('photo-disk') and inner de Vaucouleurs ('photo-bulge') components using the \GimTwoD\ two-dimensional model fitting code \citep{Marleau98,Simard02}. Bulges are assigned global, 'disk-free' colours coming from constrained \GimTwoD\ fits to the $F606W$ and $F814W$ images.

Our bulge selection provides 17 galaxies in common with \citet{Koo05}, or 20\% of their sample.  This low number is due in part to the more restricted range of redshifts covered by \citet{Koo05}. Moreover, they do not exclude galaxies morphologically classified as mergers from their sample while we do. However, the strongest contributor to the small overlap is the diameter criterion we imposed: only 21 of the 86 galaxies in \citet{Koo05} have diameters above 1.4\arcsec.  It looks as if many of the galaxies in \citet{Koo05} are spheroids that lack visible extended disks, whereas our diameter criterion generally ensures the presence of an extended, luminous disk. This explains some of the differences between the colour distributions obtained by \citet{Koo05} and us.

\section{Redshift distribution}
\label{sec:redshiftdistribution}

Figure~\ref{fig:viobs085_z} shows the observed $(F606W - F814W)$  colours vs redshift.  Colours will be analized in the  following sections.  Here, we focus on the redshift distributions  of the various sub-samples.  While there are differences between the $z$ distributions of the high- and low-inclination bulge samples, the overall redshift distribution of the galaxies with bulges is  approximately uniform in the range $0.2 < z < 0.8$. Beyond this  redshift, galaxies with measurable bulges become very scarce.  The  farthest object in the bulge sample belongs to the inclined sample and  lies at $\zspec =1.0235$.
In contrast, the non-bulge diameter-limited subsample extends to about  $z \sim 1.2$.  The cut at $z\sim 1.2$ is largely imposed by the  paucity of redshift information, and also by the diameter limit of  our sample  (Paper I, Sect. 3.3 and 4.1).

It is tempting to infer from Fig.~\ref{fig:viobs085_z} that bulges are truly more scarce at $z>0.8$.  However, several selection effects are at play.  
Galaxies with bulges may fall out of the bulge sample in two  situations. First, when apparent galaxy sizes become small to the  point that galaxies harboring bulges do not fulfill the $R>1.4$\arcsec\ size criterion.
Galaxies with bulges will fail the diameter condition at lower redshifts than non-bulge galaxies if their disks are redder  than those of non-bulge galaxies, because redder populations have stronger K-corrections in the $F814W$ band, yielding a more pronounced $z$-decrease  of $F814W$-band observed surface brightness than galaxies without  bulges.  We show below that disks of galaxies with bulges are indeed  redder than those of galaxies without bulges.

Bulges may also fall out of the sample if the surface brightness contrast between bulges and their surrounding disks becomes less prominent, leading to bulge galaxies being listed with the galaxies without bulges. The surface brightness contrast between bulges and disks, and its $z$-evolution, are affected by  both K-corrections and evolutionary corrections. K-correction effects are likely to be small, given that, for most galaxies in our sample, colour profiles are shallow (Sect.~\ref{sec:colourprofiles}). However, in principle, any galaxy with negative colour gradients suffers stronger K-corrections in the redder central parts; hence, the center regions become dimmer  with $z$ faster than the outer parts, yielding a loss of surface brightness contrast between bulges and disks as we look at higher $z$. The effects of evolutionary corrections on the bulge-disk contrast are more difficult to estimate, as they depend on the model adopted for the growth of each of the two components.

\begin{figure}[!htbp]
\begin{center}
\includegraphics[angle=0,width=0.45\textwidth]{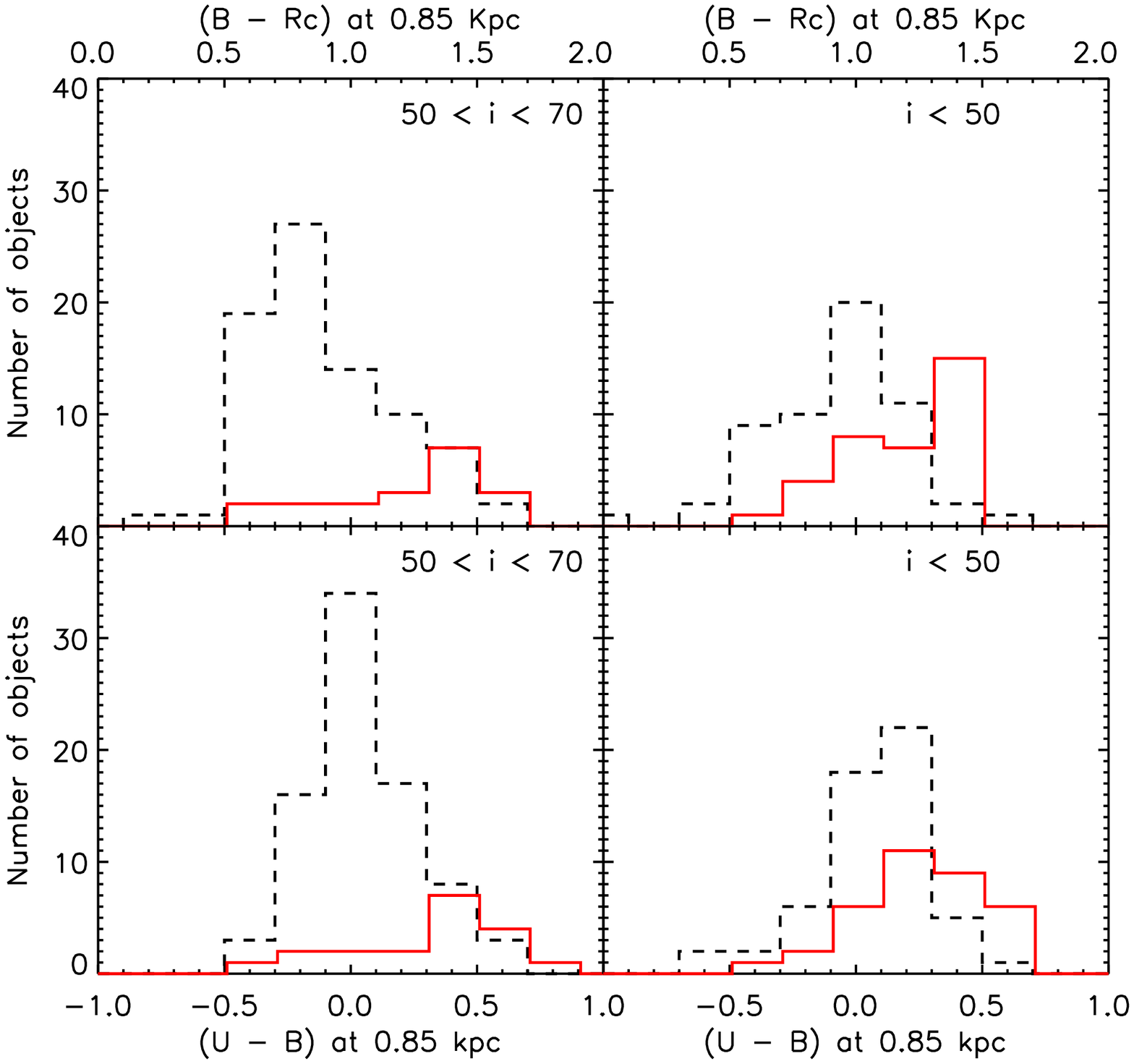}%
\end{center}
\caption{Histograms for the rest-frame central $(B - R)$ and $(U-B)$ colours measured at 0.85 kpc from the center along the minor axis for the high- (left panels) and low-inclination (right panels) samples. {\it Solid line} (red in electronic edition): colours of galaxies with prominent bulges ($\eta > 1$). {\it Dashed line}: colours of galaxies without bulges ($\eta < 1$). }   \label{fig:hist_br085}
\end{figure}

The loss of spatial resolution with $z$ may also contribute to  galaxies with bulges falling out of the sample at high $z$ because of the loss of concentration in the profiles. This  process is unlikely to be responsible for the loss of bulges at $z>0.8 $, given that loss of spatial resolution is low between $z=0.8$ and  $z=1.2$ for the concordance cosmology \citep[e.g.,][]{Weinberg72},  but must contribute to the loss of small bulges at lower redshifts. 

We carried out simulations aimed at quantifying the effects related  to K-corrections.  Each of the galaxy images in our sample was  shifted to simulate how it would appear in the GSS if the galaxy were  at higher redshift, applying repixelation, cosmological dimming and K-corrections derived from best-fit SEDs. Each image was then analysed  with \SExtractor\ to determine the maximum redshift \zmax\ at which  it would make it into the sample.  We found that galaxies with bulges retained their two-component, bulge-disk surface brightness profile to their \zmax, although bulge profiles are less concentrated with $z$, as a consequence of the effects mentioned above. This suggests  that the observed paucity of bulges beyond $z \sim 0.8$ is probably  apparent. Galaxies with bulges may exist at higher redshifts,  but they do not fulfill the diameter and central brightness excess selections of the parent sample, due to the cosmological fading of their disks and to the loss of concentration of their surface brightness profiles.

This conclusion is in agreement with the study of \citet{Schulz03},  who analysed the variations of bulge-to-disk ratios with $z$ under  for three standard bulge-disk formation scenarios, namely, bulges and  disks of equal age; old bulges and delayed disk star formation; and, old disks with subsequent bulge star formation.  After taking into account both evolutionary and band-shifting effects, these authors  find that $I$-band apparent bulge-to-total ratios ($B/T$) \textit {increase} up to $z\sim 0.9$ as a result of the combined effects of  evolution and pass-band shifting.  The reason is the progressive loss  of the disk in the noise. In our own study of bulge colours in the  deeper GOODS images, the redshift distribution of galaxies with  bulges extends well beyond $z=1$.

\begin{figure}[!htbp]
\begin{center}
\includegraphics[angle=0,width=0.45\textwidth]{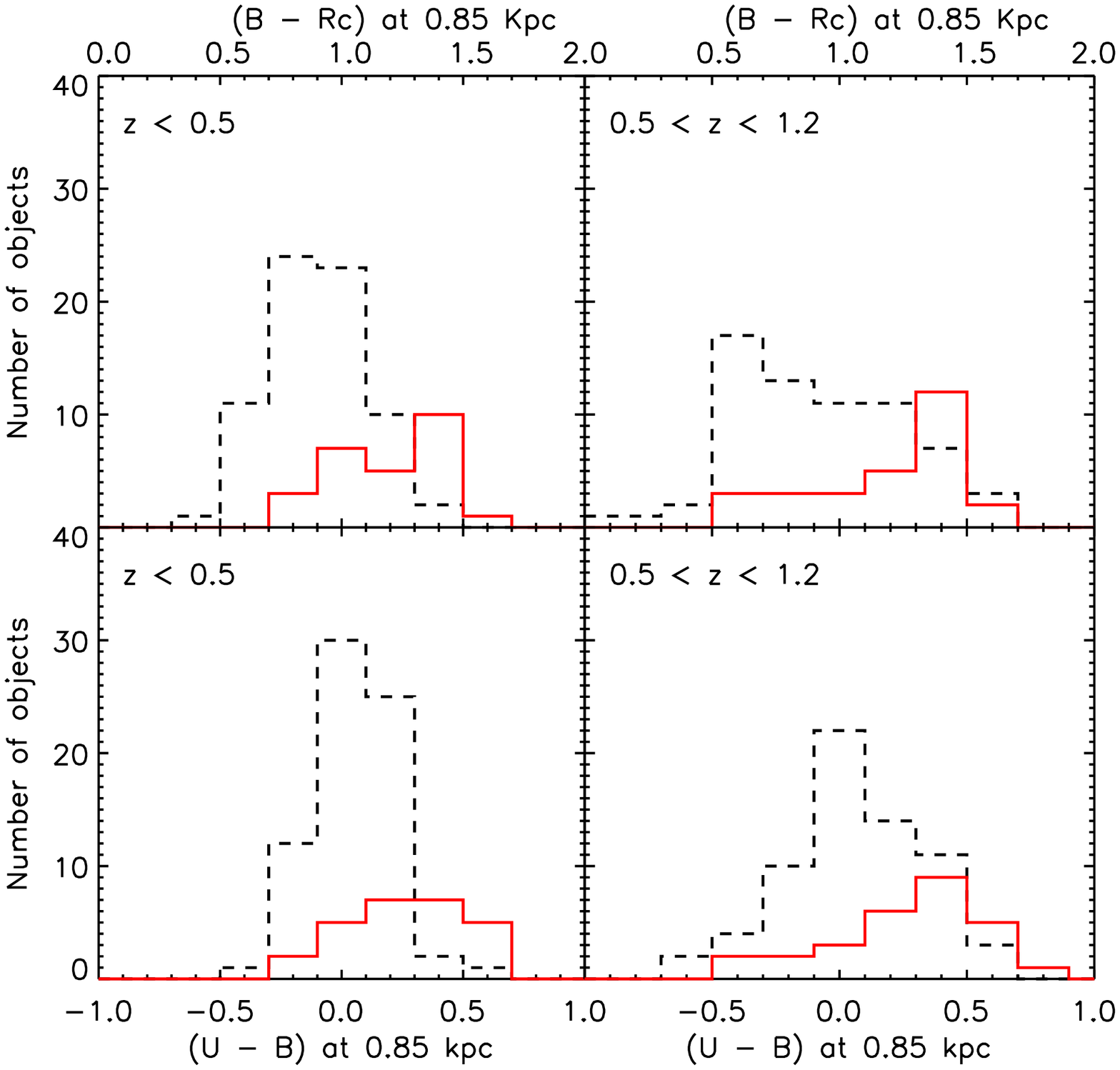}%
\end{center}
\caption{Histograms for the rest-frame central $(B - R)$ and $(U-B)$ colours measured at 0.85 kpc from the center along the minor axis for the subsample with $z<0.5$ ({\it left}) and the subsample with $0.5 < z <1.2$ ({\it right}). {\it Solid line} (red in electronic edition): colours of galaxies with prominent bulges ($\eta > 1$). {\it Dashed line}: colours of galaxies without bulges ($\eta < 1$).} \label{fig:hist_ub_tot_0512}
\end{figure}

\section{Colour profiles and colour gradients}
\label{sec:colourprofiles}

The $(F606W-F814W)$ colour profiles for both semi-minor axes of our samples are shown in Paper I (Appendix B). We see that, for most galaxies, one profile is bluer than the other. The bluer profile is the one we use to calculate the nuclear colours (Paper I, Sects. 5.1 and 5.2). Colour profiles for the bluer semi-minor axes, geometrically deprojected to face-on, are provided in Appendix C of Paper I.

For the bulge sample, inspection of the colour profiles in Appendix~C (Paper~I) shows that most colour profiles are smooth and show gentle gradients, bluer outward.  Only in four out of the 54 bulge galaxies do we find nuclear positive colour gradients, which probably reflect an internal structure such as a disk, ring or bar harboring star formation. None shows a sudden inner reddening at the region of the bulge.
Colour profiles give important clues on the colour differences between bulges and disks. Indeed, the smoothness of the colour profiles, as we go from bulge-dominated to disk-dominated radii, indicates that the bulge-disk structure is not responsible for the colour gradient. 

For the non-bulge galaxies, inspection of the profiles in Appendix~C (Paper~I) reveals a wider range of profile shapes and slopes, but profiles tend to show a gentle  negative gradient, bluer outward.  To provide a quantitative comparison of bulge and non-bulge colour profiles, colour gradients are shown in Fig.~\ref{fig:grad_eta} against the central brightness prominence $\eta$ (Sect.~\ref{sec:data}).  Gradients, computed from linear fits to the colour profiles, are shown separately for the bluer semi-minor axis; for the redder semi-minor axis; and, for the major axis.  
If colour gradients were driven by a central redder bulge surrounded by a bluer disk, gradients would correlate with $\eta$.  Figure~\ref{fig:grad_eta} shows that in no case a correlation exists between colour gradients and $\eta$: a higher central brightness prominence does not imply a stronger gradient. In Table~\ref{tab:colgradient} we show median values and the standard deviations of colour gradients for low- and high-inclination samples and for bulge and non-bulge samples. 
For bulge galaxies, the high-inclination sample has, in general, larger negative gradients than the low-inclination one; see in particular the gradients corresponding to the bluer semi-minor axes. This may be explained by the effect of residual dust reddening in the center in inclined galaxies (see Sect. 2, Paper I), which leads to stronger negative gradients in the bluest side of the galaxy. In any case, we see that the colour gradients are rather small in all cases and, are slightly stronger in the non-bulge samples. This again shows that the bulge-disk structure is not responsible for the colour gradients.

\begin{table*}
\caption{Median $(B-R)$ colours and dispersions} \label{tab:statBR}
\begin{center}
\begin{tabular}{l c c c c c c c c}
\hline
\hline
  & \multicolumn{4}{c}{$i < 50$} & \multicolumn{4}{c}{$50 < i < 70$}\\
\hline
 & \multicolumn{2}{c}{$\eta > 1$} & \multicolumn{2}{c}{$\eta < 1$} & \multicolumn{2}{c}{$\eta > 1$} & \multicolumn{2}{c}{$\eta < 1$}  \\
\hline
 & med & $\sigma$ & med & $\sigma$ & med & $\sigma$ & med & $\sigma$ \\
\hline
 0.85 kpc & 1.24 & 0.23 & 0.97 & 0.29 & 1.30 & 0.33 & 0.86 & 0.28 \\
 galaxy & 1.19 & 0.24 & 0.87 & 0.18 & 1.13 & 0.21 & 0.81 & 0.25 \\  
 disk & 1.19 & 0.25 & 0.88 & 0.26 & 1.03 & 0.23 & 0.83 & 0.29 \\
\hline
\hline
\end{tabular}
\end{center}
\begin{footnotetext}
TNote.- $(B-R)$ median colours and standard deviation for samples with bulges ($\eta > 1$) and without bulges ($\eta < 1$) and for the low- ($i < 50$\deg) and high- ($50\deg < i < 70$\deg) inclination samples
\end{footnotetext}
\end{table*}

\section{Observer-frame nuclear colours}
\label{sec:observedcolours}

In Fig.~\ref{fig:viobs085_z} we show the observed $(F606W - F814W)$ colours vs redshift for both the high- (left panel) and low-inclination (right panel) samples. 
At all redshifts, bulges tend to cluster along the red envelope of the distribution.  This red envelope is well described by passively evolving populations (SSP models with solar metallicity, \citet{Chabrier03} IMF, no internal dust).  Bulges along the red envelope are not forming stars at the observation epoch.  Large changes in  the formation redshift of the models ($\zf = 10, 3$, and 1.5 are shown in the figure) yield similarly good fits: the red envelope provides null constraints on the formation redshift.  

Figure~\ref{fig:viobs085_z} also shows that a notable fraction of all bulges lie well below the red envelope. We quantify this statement below, when we present rest-frame colours. A similar colour dispersion was found by \citet{Ellis01}.   These authors proposed a model in which minor episodic star formation events would make old bulges temporarily populate the bluer regions of the diagram.  A scenario with episodic star formation was also suggested by \citet{Koo05}, for a different reason, namely, that the virtual equality of  rest-frame bulge colours at $z=0.8$ and $z=0$ is inconsistent with passive evolution.  We discuss formation scenarios in Sect.~\ref{sec:discussion}.

The distribution of nuclear colours for non-bulge galaxies ($\eta < 1$) is markedly different from that of the galaxies with bulges. Non bulge galaxies are dominantly bluer. While some of them may reach up to the bulge red envelope, above all in the high-inclination sample, they do not cluster at redder colours as bulge galaxies do.

\begin{table*}
\caption{Median $(U-B)$ colours and dispersions} \label{tab:statUB}
\begin{center}
\begin{tabular}{l c c c c c c c c}
\hline
\hline
  & \multicolumn{4}{c}{$i < 50$} & \multicolumn{4}{c}{$50 < i < 70$}\\
\hline
 & \multicolumn{2}{c}{$\eta > 1$} & \multicolumn{2}{c}{$\eta < 1$} & \multicolumn{2}{c}{$\eta > 1$} & \multicolumn{2}{c}{$\eta < 1$}  \\
\hline
 & med & $\sigma$ & med & $\sigma$ & med & $\sigma$ & med & $\sigma$ \\
\hline
 0.85 kpc & 0.28 & 0.23 & 0.11 & 0.24 & 0.38 & 0.32 & 0.06 & 0.20 \\
 galaxy & 0.27 & 0.24 & -0.04 & 0.16 & 0.29 & 0.22 & -0.03 & 0.17 \\  
 disk & 0.23 & 0.25 & 0.00 & 0.24 & 0.17 & 0.25 & -0.01 & 0.22 \\
\hline
\hline
\end{tabular}
\end{center}
\begin{footnotetext}
TNote.- $(U-B)$ median colours and standard deviations for samples with bulges ($\eta > 1$) and without bulges ($\eta < 1$) and for the low- ($i < 50$\deg) and high- ($50\deg < i < 70$\deg) inclination samples
\end{footnotetext}
\end{table*}

\section{Rest-frame nuclear colours}
\label{sec:restframecolours}

In Fig.~\ref{fig:br_z} we show rest-frame nuclear colours vs redshift. Left and right panels show the high- and low-inclination samples, respectively, and upper and lower panels the rest-frame $(B-R)$ and $(U-B)$ colours, respectively. This figure is the K-corrected version of Fig.~\ref{fig:viobs085_z}, and emphasizes the clustering of bulge galaxies on the red sequence of passively evolving populations ($B - R \sim 1.3 - 1.5$ and $U - B \sim 0.3 - 0.6$), over the $0.2 < z < 0.9$ range where we detect bulges. In the high-inclination sample, three bulges and two non-bulges lie more than 0.1 mag above the passive evolution line; we suspect this is due to residual dust reddening.  Blue bulges exist as well at all redshifts.  
Colours of galaxies with low central prominence ($\eta < 1$) uniformly populate the entire region $0.5 < B-R < 1.3$, $-0.5 < U-B < 0.5$, typical for star-forming populations, and they show no concentration toward red colours. 

Because the red envelopes of the distributions in Fig.~\ref{fig:br_z} are close to horizontal, we collapse the $(B - R)$ and $(U - B)$  colour distributions into histograms (Fig.~\ref{fig:hist_br085}).   Bulge samples are dominately redder than the non-bulge samples, in  all of the plots. Median and rms values of the $(B-R)$ and $(U-B)$  colour distributions are listed in Tables~\ref{tab:statBR} and \ref {tab:statUB}, respectively: 50\% of the bulge sample ($\eta > 1$)  shows bulge colours $(B-R) > 1.27$ and $(U-B) > 0.33$ (after averaging  low- and high-inclination values).  The fraction of 'very red bulges'  \citep[$U-B > 0.25$ in the terminology of ][]{Koo05} is 60\%.
In contrast, median values for the low-prominence ($\eta < 1$)  galaxies are $(B-R) = 0.91$ and $(U-B) = 0.08$.

We use a K-S test to inquire whether the various samples come from  similar or different parent distributions.  The results, given in  Table~\ref{tab:KST}, show that we reject the hypothesis that colours  of bulges and non-bulges come from the same parent distribution with  over 99.99\% confidence (99.80\% when samples are split by  inclination).  Conversely, Table~\ref{tab:KST} shows that, in regards  their nuclear colours, high- and low-inclination samples of bulges  come from the same parent distribution, and the same occurs for the  high- and low-inclination sample of non-bulge galaxies.  
 
Recalling that the bulge sample has  been defined only from the central brightness prominence, $\eta$, the histograms in Fig.~\ref{fig:hist_br085} indicate that, for the redshifts covered by our data, galaxies with positive central deviations larger than 1 magnitude with respect to an exponential profile show a strong tendency toward red population colours. Hence, the association of red colours with central concentration, well known in the local Universe \citep{Hogg04}, is also found at redshifts up to $z=0.9$.

\begin{table*}
\caption{Kolmogorov-Smirnov tests} \label{tab:KST}
\begin{center}
\begin{tabular}{c c c c c c}
\hline
\hline
  & H.L bul & H.L no-bul & Hbul.Hno-bul & Lbul.Lno-bul & bul.no-bul \\
  & (1) & (2) & (3) & (4) & (5)\\
\hline
$(B-R)$ & 6.3 & 4.1 & $2.9\times10^{-2}$ & $2.0\times10^{-1}$ & $9.8\times10^{-5}$ \\
$(U-B)$ & 32.5 & 48.6 & $4.1\times10^{-3}$ & $2.5\times10^{-2}$ & $2.9\times10^{-6}$ \\
\hline
\hline
\end{tabular}
\end{center}
\begin{footnotetext}
TNote.- Percentage probability to belong to the parent distribution:
(1) High-inclined bulges  vs low-inclined bulges 
(2) High-inclined non-bulges vs low-inclined non-bulges 
(3) High-inclined bulges vs high-inclined non-bulges 
(4) Low-inclined bulges vs low-inclined non-bulges 
(5) Bulges vs non-bulges (high-inclination + low-inclination)
\end{footnotetext}
\end{table*}

We note here that selecting bulges in our initial sample by DEEP bulge magnitude ($F814W > 23.57$ in Vega system), does not lead to a so well defined colour bi-modality between bulge and non-bulge galaxies; the nuclear colour distributions show median values: $(B-R) = 1.09$ and $(U-B) = 0.16$ for galaxies with bulges and $(B-R) = 0.76$ and $(U-B) = 0.02$ for galaxies without bulges.  We get bluer bulges because, from our inspection of the images and the surface brightness profiles, we find that the  galaxies selected by DEEP bulge magnitude include galaxies with nearly pure exponential profiles, as well as mergers (see Fig. 5 from Paper I). 

We have investigated possible redshift trends of the colour distribution of bulges.  When splitting at $z=0.5$, we note a marginal tendency for flatter bulge colour distributions at $z<0.5$, and a more pronounced red peak at $z>0.5$ (Fig.~\ref{fig:hist_ub_tot_0512} ):  we find more blue bulges at $z<0.5$ than at higher redshift.  This is contrary to the expectation that the bulge formation epoch would be found at high $z$. The result is due to the higher spatial resolution at low $z$, which allows us to detect smaller central light prominences - small bulges in formation. In any case, Fig.~\ref{fig:hist_ub_tot_0512} shows that the colour segregation between bulges and non-bulges persists when the sample is split in two redshift ranges.  

\begin{figure*}
\begin{center}
\includegraphics[angle=0,width=0.9\textwidth]{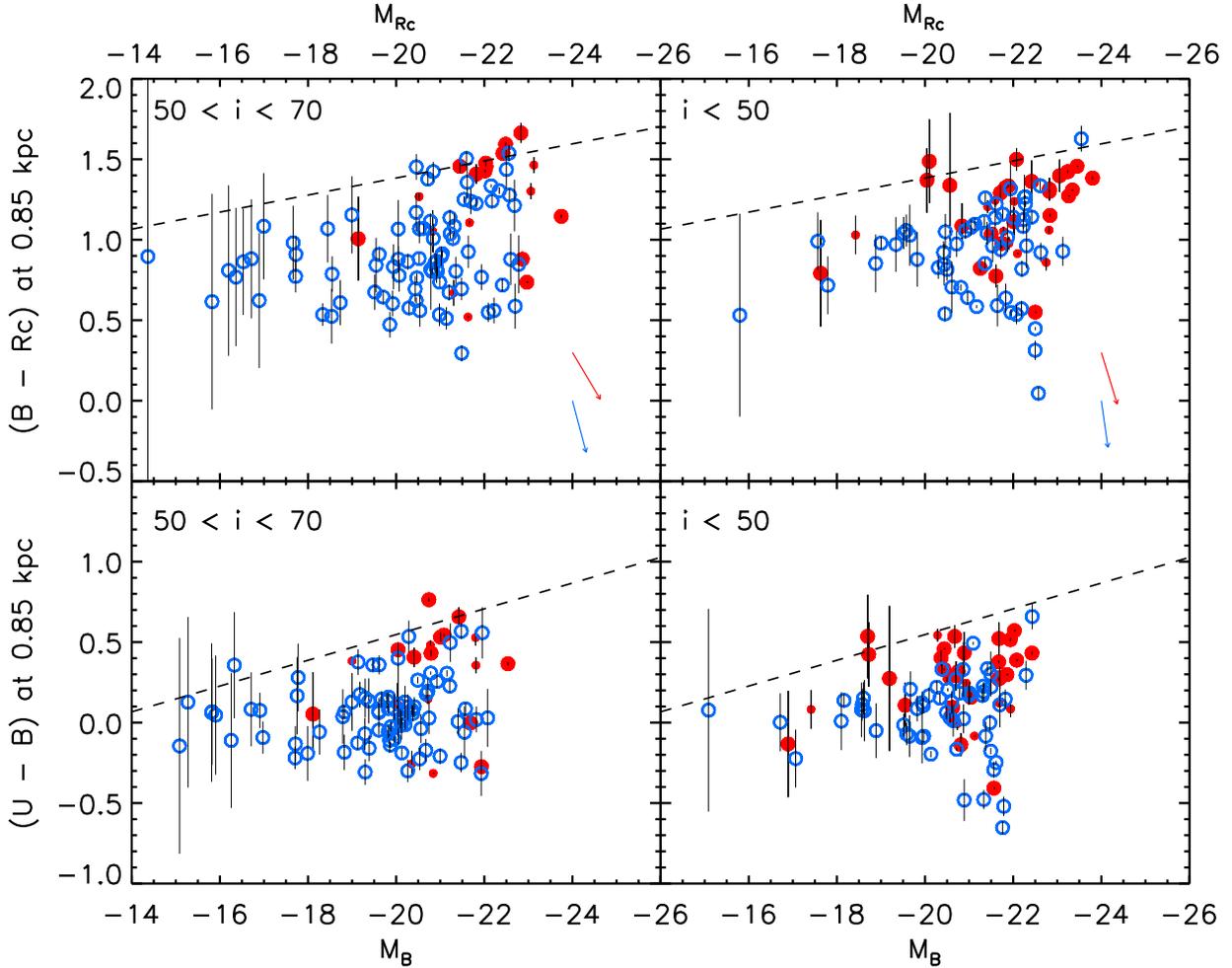}%
\end{center}
\caption{Colour-magnitude diagrams: $(B-R)$ vs $M_R$, $(U-B)$ vs $M_B$ for the high- (left panels) and low-inclination (right panels) samples. {\it Filled circles} (red in the electronic edition): colours of galaxies with prominent bulge ($\eta > 1$). {\it Smaller filled circles}: galaxies with prominence indices in the range $1 < \eta < 1.5$. {\it Open circles} (blue in the electronic edition): colours of galaxies without bulges ($\eta < 1$). {\it Dashed line}: colour-magnitude distribution for early-type galaxies in the Coma cluster, derived from the CMDs of \citet{Eisenhardt07} and \citet{Scodeggio01}.  {\it Red vectors}: reddening-attenuation vectors for galaxies with $B/T=0.4$, central $B$-band opacity $\tau_B = 4$, and inclinations of $60\deg$ and $30\deg$, respectively. {\it Blue vectors}: reddening-attenuation vectors for pure disk galaxies with central $B$-band opacity $\tau_B = 4$, and inclinations of $60\deg$ and $30\deg$, respectively.} \label{fig:ub_bmag}
\end{figure*}

In Fig.~\ref{fig:ub_bmag} we show colour-magnitude diagrams: $(B - R)$ vs $M_R$ and $(U - B)$ vs $M_B$, in the same format as Fig.~\ref{fig:br_z}. We plot nuclear colours measured at 0.85 kpc vs the total absolute magnitude of the galaxy.
The dashed line is the colour-magnitude diagram (CMD) for early-type galaxies in the Coma cluster, derived from the CMDs of \citet{Eisenhardt07} and \citet{Scodeggio01}. 
Vectors, indicating the corrections that need to be applied to the data points to take into account the effects of dust, are plotted for the extreme case of a thin, opaque dust layer in the mid plane of the galaxies. Using \citet{Tuffs04} and \citet{Moellenhoff06}, we get, for galaxies with $B/T=0.4$ and with a central $B$-band disk opacity $\tau_B = 4$, attenuations of $\sim0.9$, and $\sim0.6$ mag for inclinations of $60\deg$ and $30\deg$, respectively, for $M_B$. The atenuations for $M_R$ are $\sim0.6$ and $\sim0.36$ for the same inclinations. The corresponding reddenings in $(B-R)$, for nuclear colours derived from surface brightness profiles, are $\sim0.29$ and $\sim0.32$. For pure-disk galaxies, we obtain, for the same opacity and inclinations, attenuations of the order of $\sim0.46$ and $\sim0.26$ for $M_B$ and $\sim0.3$ and $\sim0.15$ for $M_R$, reddening in $(B-R)$ are $\sim0.26$ and $\sim0.29$. In this model, reddenings are moderate because, in the limit of the opaque disks, we only see the unreddened near side.

The bulge distributions show two salient features. First, the reddest bulges, which are almost as red as local early-type galaxies if extinction is null, would appear well below the colour-magnitude relation of local early-type galaxies if extinction as estimated in the previous paragraphs were assumed, yielding stellar populations with comparatively lower ages and/or metallicities. And second, we find that the absolute magnitudes are constrained to the bright range, approx $\sim -20 < \MR < -24$ and $\sim -19 < \MB  < -23$ without dust attenuation correction, for the whole range of bulge colours. 
The lack of bulges in low luminosity galaxies is not due to a selection effect, as any low-luminosity galaxies with bulges would have been seen by our selection criteria.

In contrast, absolute magnitudes of low-prominence galaxies ($\eta < 1$) lie in a broader range: $-16 < \MR < -23$ and $-15 < \MB < -22$, without dust attenuation correction. The percentage of faint galaxies is larger in the high-inclination sample than in the low-inclination one, we get that $33\%$ of low-prominence galaxies in the high-inclination sample are fainter than $\MR \sim -20$ and $\MB \sim -19.5$, while this percentage decreases to $18\%$ in the low-inclination one. This is a selection effect, as high-inclination galaxies fulfill the $1.4$\arcsec\ size criterium to fainter limits.

\begin{figure*}
\begin{center}
\includegraphics[angle=0,width=0.9\textwidth]{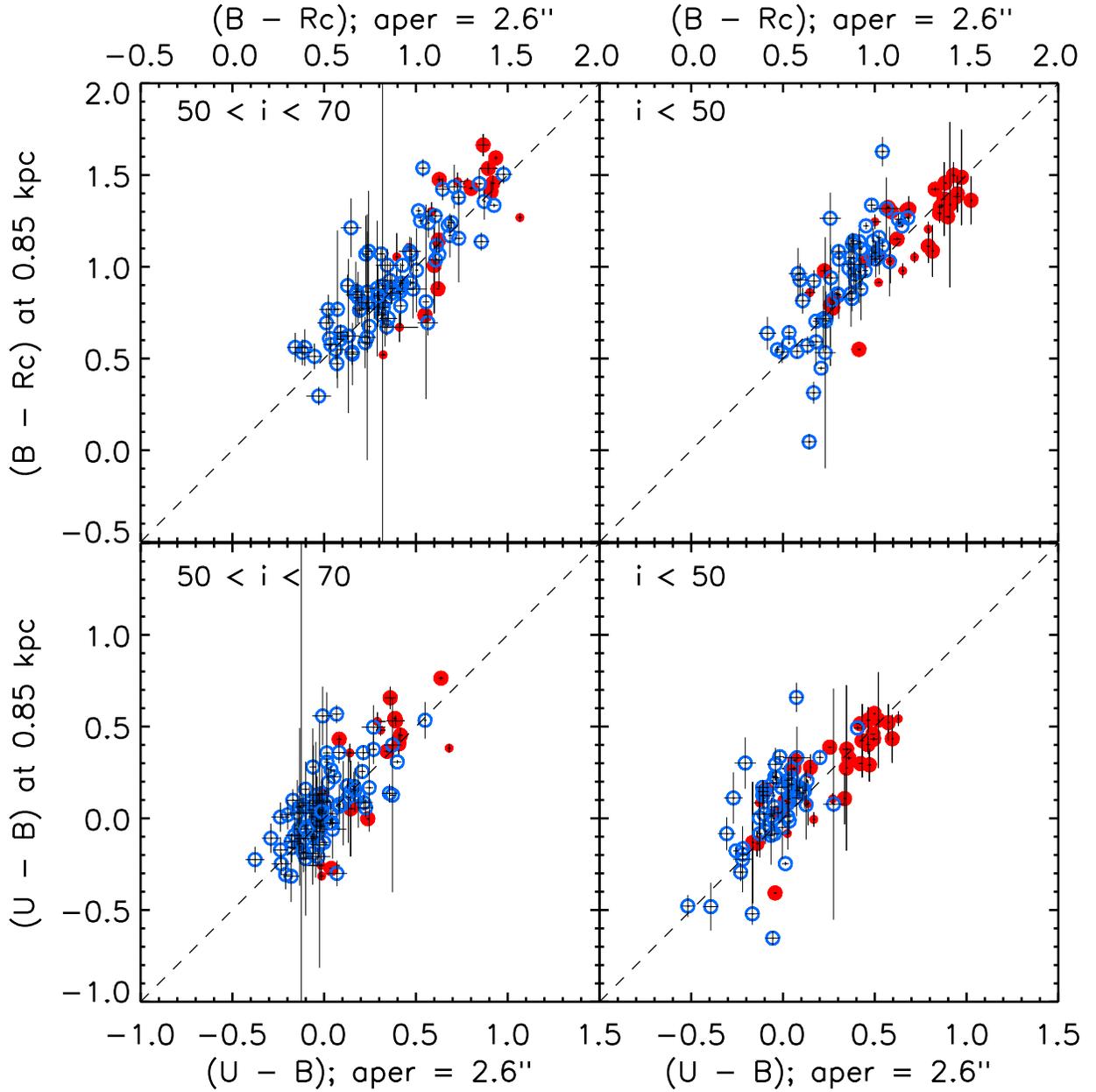}%
\end{center}
\caption{Rest-frame $(B-R)$ and $(U-B)$ colours measured at 0.85 kpc vs the integrated rest-frame colour of total galaxy for the high- (left panels) and low-inclination (right panels) samples. Integrated colours are measured in a $2.6''$ diameter aperture. {\it Filled circles} (red in the electronic edition): colours of galaxies with prominent bulge ($\eta > 1$). {\it Smaller filled circles}: galaxies with prominence indices in the range $1 < \eta < 1.5$. {\it Open circles} (blue in the electronic edition): colours of galaxies without bulges ($\eta < 1$).}  \label{fig:colrestaper2_colrestbul}
\end{figure*}

\section{Rest-frame global colours and disk colours}
\label{sec:restframeglobalcolours}

How nuclear colours relate to galaxy global colours can be seen in Tables~\ref{tab:statBR} and~\ref{tab:statUB} and in Fig.~\ref{fig:colrestaper2_colrestbul}, which shows rest-frame ($B - R$) and ($U - B$) at 0.85 kpc vs the rest-frame ($B - R$) and ($U - B$) colours measured in a $2.6''$ diameter aperture. Nuclear colours and global colours are very similar: nuclei are redder than their parent galaxies by $\Delta(B-R) \sim 0.05$, averaging all objects in all samples, whereas colours of nuclei span a range of over 1 mag in both $(B-R)$ and $(U-B)$.  

\begin{figure*}
\begin{center}
\includegraphics[angle=0,width=0.9\textwidth]{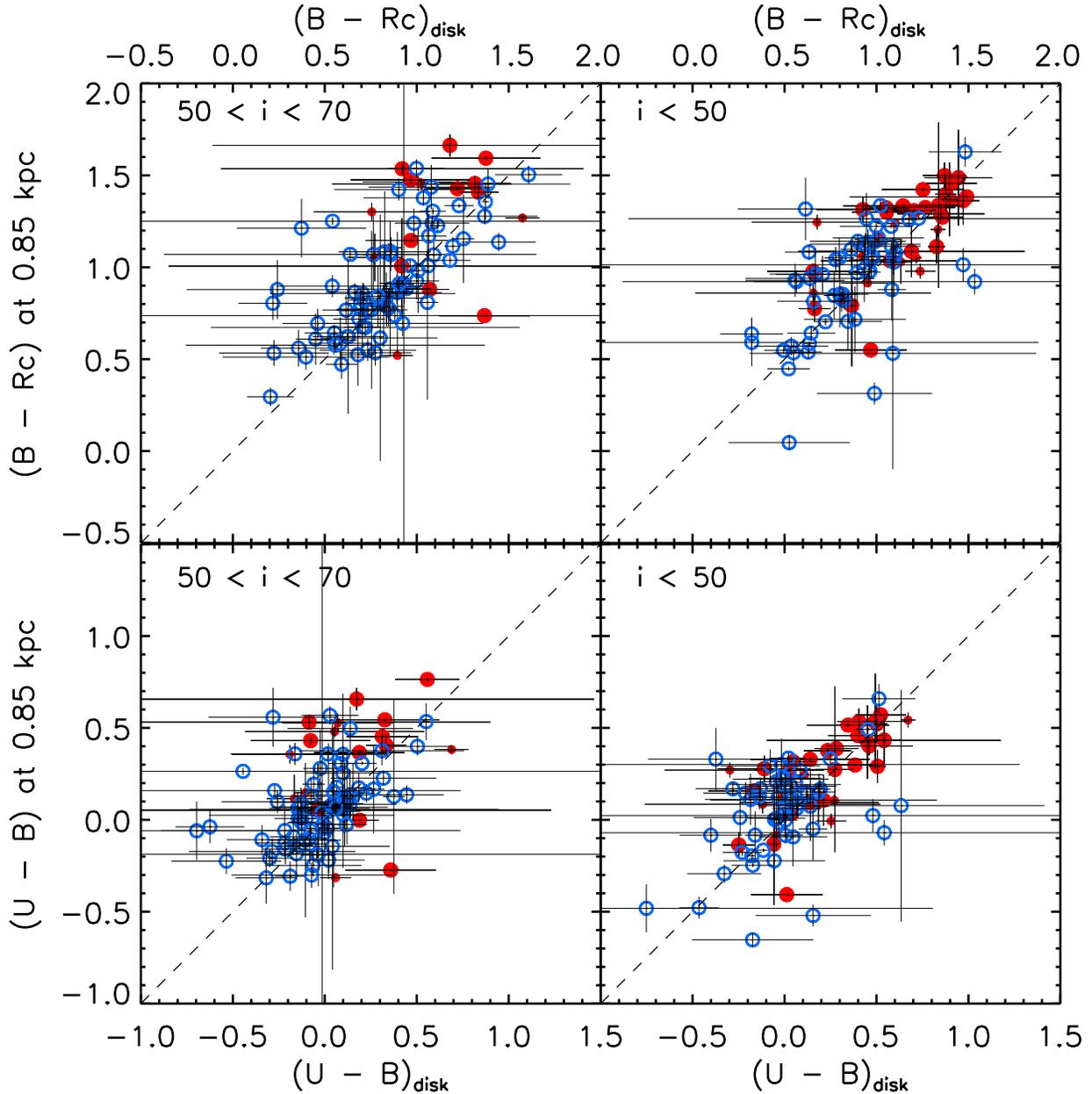}%
\end{center}
\caption{Rest-frame $(B-R)$ and $(U-B)$ colours measured at 0.85 kpc vs the rest-frame colour of the disk for the high- (left panels) and low-inclination (right panels) samples measured at 0.85 kpc. {\it Filled circles} (red in the electronic edition): colours of galaxies with prominent bulge ($\eta > 1$). {\it Smaller filled circles}: galaxies with prominence indices in the range $1 < \eta < 1.5$. {\it Open circles} (blue in the electronic edition): colours of galaxies without bulges ($\eta < 1$).}  \label{fig:colrestdisk_colrestbul}
\end{figure*}

One may argue that the correlation between bulge and total colours might be driven by the fact that the bulge light contributes to the total galaxy colour.  However, the similarities remain when we compare bulge colours to disk colours (Tables~\ref{tab:statBR} and~\ref{tab:statUB}, and Fig.~\ref{fig:colrestdisk_colrestbul}; the methodology for the derivation of disk colours is detailed in Paper I, Sect. 5.3).  The trend is weaker, but it is nevertheless present.  
The correlation is stronger in the low-inclination sample than in the high-inclination one.  In the latter, we find instances of disks that are much bluer than their nuclei. The larger error bars in the colours of the inclined disks probably account for the weaker correlation; particularly, in the sample of galaxies with $\eta > 1$, this happens for 2 galaxies (with id numbers: 8, 59; see Paper I, Table A.3), which are very faint ($I_{814} > 21.3$) and have very noisy disks.  Residual dust reddening of the inclined bulge colours cannot be ruled out either. Four galaxies in the sample with $\eta > 1$ (with id numbers: 162, 116, 150, 129) show distorted and noisy disks, which usually are associated to star formation and may contribute to making nuclei more dusty. In the high-inclination sample we also find a high-prominence galaxy with a nucleus much bluer than its disk (id number: 142), this galaxy also has a distorted disk and shows a possitive colour gradient.

We gather from those figures that most galaxies with redder bulges have redder disks; the colour difference between bulges and disks is smaller than that between bulges of different galaxies. 
We believe that disk contamination does not drive the observed coupling of disk and bulge colours. As shown in Paper~I (Sect.~6.1), if the disks have uniform colours then the bulge intrinsic colours might be redder than our measurement by an average of $\Delta(B-R) = 0.13(0.03)$ for the high- (low-)inclination samples.  And, because most disks do not have uniform colours but show negative colour gradients, disk contamination must in general be smaller than these estimates. The corrections are higher for the high-inclination sample because the disks become brighter at higher inclinations. But dust may contribute to the observed similarities of disk and bulge colours, by hiding young populations.  

\begin{figure*}
\begin{center}
\includegraphics[angle=0,width=0.9\textwidth]{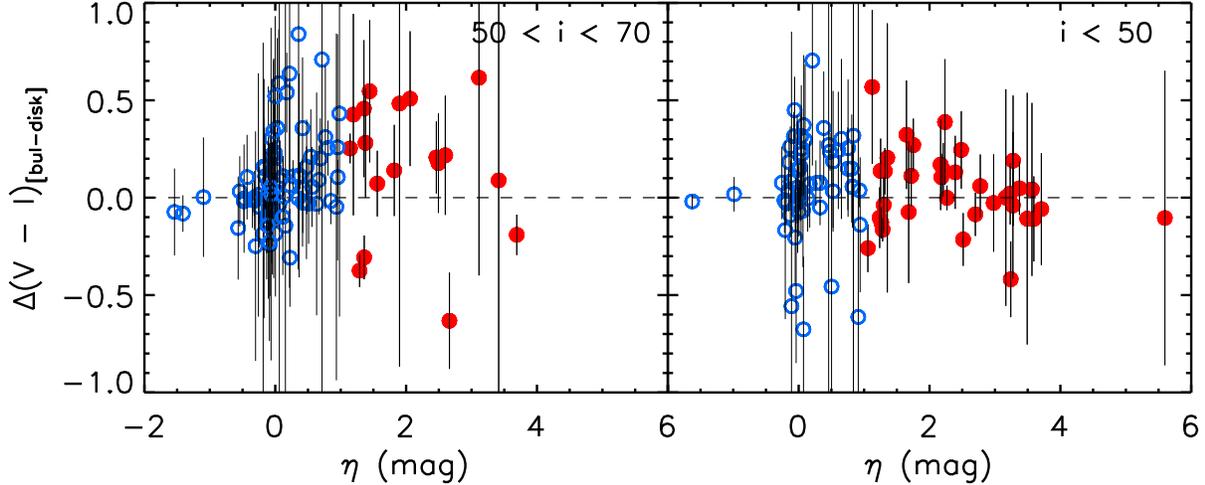}%
\end{center}
\caption{Differences between bulge colour at 0.85 kpc and disk colour vs the central brightnes excess, $\eta$ for the high- (left panel) and low-inclination (right panel) samples measured at 0.85 kpc. {\it Filled circles} (red in the electronic edition): colours of galaxies with prominent bulge ($\eta > 1$). {\it Open circles} (blue in the electronic edition): colours of galaxies without bulges ($\eta < 1$).}  \label{fig:viobsdif_viobsexcess}
\end{figure*}

The differences between $(F606W-F814W)$ colours observed at 0.85 kpc  and in the disks are shown in Fig.~\ref{fig:viobsdif_viobsexcess} vs the central brightness prominence $\eta$.  This figure may be compared to Fig.~\ref{fig:grad_eta}, as both portrait similar information.  The colour difference does not correlate with central brightness prominence; bulges and disks have similar colours differences no matter how concentrated is the surface brightness profile. Median value and standard deviation of the colour differences for each sub-sample are shown in Table~\ref{tab:stdevdifbuldisk}.  The median difference for the entire sample is very small, $\Delta(V-I) = 0.05$.  Only for the inclined bulges does this difference increase, to a moderate $\Delta(V-I) = 0.21$.

\begin{table}
\caption{Bulge-disk colour differences: $(V-I)_{0.85}-(V-I)_{disk}$} \label{tab:stdevdifbuldisk}
\begin{center}
\begin{tabular}{l c c c c}
\hline
\hline
  & \multicolumn{2}{c}{$i<50$} &  \multicolumn{2}{c}{$50<i<70$} \\ 
\hline
   & med & $\sigma$ & med & $\sigma$ \\
\hline
all $\eta$ & 0.05   &  0.23  & 0.05 & 0.25 \\
$\eta >1$ & 0.02   &  0.19 & 0.21 & 0.34  \\
$\eta <1$ & 0.07   &  0.26 & 0.03 & 0.22  \\
\hline
\hline
\end{tabular}
\end{center}
\end{table}

\section{Discussion}
\label{sec:discussion}

We have seen that, in a diameter-limited sample of disk galaxies ($0.1 < z < 1.3$), a surface brightness profile with an inner excess with respect to the outer exponential is most often associated with colours typical of the red sequence; inasmuch as inner surface brightness excess (referred to as prominence in this paper) traces the presence of a bulge, nominally we find 60\% 'very red' bulges ($U-B > 0.25$), and 40\% blue bulges. Our fraction of red-sequence bulges is thus lower than, the 85\% fraction of 'very red bulges' found by \citet{Koo05}'s study of bulge colours in the Groth strip.  Given the uncertainties in both studies, ranging from ambiguities in sample selection to the difficulty of assigning a given representative colour to the bulge, we do not want to put much weight in the differences in red bulge fraction in both studies.  
\citet{Koo05}'s colours are probably biased to the red, from the \RdeV-plus-exponential modeling and/or the assumption of uniform colours for bulges and disks (see Paper~I, Sect. 6.1).  But our choice of the bluest wedge, may plausibly introduce biases of our colours to the blue, if the wedge hits a patch of star formation, or simply due to rms fluctuations  (Paper~I, Sect. 5.1).

Our red bulges link with local bulges along passive evolutionary tracks (Sect.~\ref{sec:restframecolours}; Fig.~\ref{fig:br_z}).  \citet{Koo05}'s conclusion that $z\sim0.8$ bulges have the same colours as $z=0$ bulges, may be traced to their redder colours ($\Delta(V-I)\sim 0.3$; see Paper I, Sect. 6.1), which, once translated into rest-frame $(U-B)$, amounts to higher reddening than that expected from passive evolution from $z=0.8$ to $z=0$. Rejuvenation of the bulge populations is needed to link \citet{Koo05}'s measurements to local colours. In our data, we do find a few bulges with colours above the passive evolutionary tracks; these would be candidates for very old and metal-rich populations that need rejuvenation to link with $z=0$ colours; but, given that they are only found at high-inclination, a likely possibility is that their colours might be affected by residual dust reddening. On the whole, our red bulges are broadly consistent with linking to $z=0$ bulges through passive evolution of old populations.  We therefore tentatively conclude that there is little evidence for rejuvenation in the colours of our red bulges. However, given the small sample sizes and the small reddening implied in passive evolution from $z=0.8$ to $z=0$ (see Fig.~\ref{fig:br_z}), whether red bulges at intermediate-$z$ do travel back and forth between the red sequence and the blue cloud remains an open question.  

In the debate on what fraction of bulges/spheroids are undergoing active star formation (Sect.~\ref{sec:introduction}), our data places us in an intermediate position between \citet{Ellis01} and \citet{Koo05}.  Blue central prominences identify star formation at the central regions of disk galaxies, that may plausibly lead to bulges after star formation ceases, hinting at the possibility that $\sim$40\% of present-day bulges have significantly grown at $z<1$. These numbers suggest a wider colour evolution of bulge populations in the past 6 Gyr than proposed by \citet{Koo05}.  \citet{Ellis01} estimated a fraction of spheroids in formation of 30\% to 50\%, which matches the fraction of blue bulges in our sample.  Note however that \citet{Ellis01}'s and our sample are very different (see Sect.~\ref{sec:introduction}), and that, contrary to them, we find few instances of positive colour gradients, a key criterion used by Ellis et al.\ to identify ongoing spheroid growth.  The latter result suggests that very few of the galaxy centers are undergoing starbursts confined to the nuclei in our disk galaxy sample; the general similarity of nuclear and global colours suggests instead that the pace of star formation in the centers and in the outer disks is not dissimilar. 

Our results also indicate that the central regions of disk galaxies, when classified by central concentration, yield a bi-modal colour distribution.  More concentrated galaxies ($\eta > 1$) tend to be redder.  Galaxies without central excess ($\eta < 1$) with respect to the outer exponential show colours typical of star forming populations, and do not cluster along the red sequence (Fig.~\ref{fig:hist_br085}). We conclude, therefore, that this bi-modality, well known in the local Universe, is already present at $z \sim 1$.  The link between density and colour has many times been explained as the result of the increasing dominance of 'the red bulge' over 'the blue disk' as we move to earlier types \citep[e.g.][]{Driver06}.  \citet{Peletier96} argue against this interpretation, citing the strong coupling of bulge and disk colours at $z=0$.  The present data supports the view that bulges and disks are closely coupled, and extend it to redshifts up to $z\sim 1$, given that, at those redshifts, bulge colours strongly scale with disk and global galaxy colours.  Colour profiles are smooth, and do not show a discontinuity at the bulge-disk boundary. Those correlations are fulfilled for all types of galaxies, independently of central concentration and colour.  

Beyond the observation that red bulges must be older than blue bulges, we now briefly address formation mechanisms. 
Because the data do not show instances of bulges surrounded by much bluer disks, we find little evidence for the 'bulge before disk' model - unless dust in disks and bulges conspires to make colours similar throughout each galaxy.  Detailed modeling would be needed to check whether the common assumption in semi-analytic models of spiral galaxy formation,  involving a major merger origin for bulges followed by the formation of a pristine new disk out of remaining gas \citep{Kauffmann96bul, Baugh96, Cole00}, is consistent with our colours.
Our data indicate that, if those processes contribute to bulge formation, they do not leave an imprint in the colours at $z\lesssim 0.8$.  
We do find four cases, out of 54 galaxies, of bulges distinctly bluer than their parent disks.  These must correspond to instances of nuclear star formation leading to bulge growth.  These galaxies do not match the central densities of luminous blue compact galaxies (LBCG) which \citet{Hammer01} propose as phases of bulge growth in disk galaxies; and their numbers are too low to account for a dominant effect in the growth of the bulge population.  At the redshifts sampled by our data, i.e., up to $z=0.8$, bulges and disks rather appear to have followed strongly intertwined star formation histories for a large fraction of their evolution.

For bulges in the blue cloud two basic models have been proposed: in a rejuvenation scenario, a red, old bulge suffers a burst of star formation, due perhaps to gas accretion, then it slowly reddens and returns to the red sequence as the burst ages.  The timescales of those processes are much longer than the collapse time, so we may be looking at the bulges in different evolutionary phases, which may be one of the reasons of the spread of colours in blue bulges. This model has been suggested by several authors  \citep[e.g.][]{Menanteau01, Ellis01, Koo05}. Alternatively, a blue nucleus might trace enhanced central star formation in the disk, perhaps related to the pseudo-bulge phenomenon in the Local Universe \citep{Kormendy04}.  Dintinguishing between these two models may be done on the basis of central surface brightness, as a starburst on top of an already dense old bulge must lead to higher surface brightness than starbursts fed by disk instabilities.  

For bulges in the red sequence, i.e., passively evolving, comparison with synthetic models (Fig.~\ref{fig:viobs085_z}) gives us clues that such bulges may have stopped forming stars at any epoch earlier than $\sim 1$ Gyr before the observation. The strong correlation of bulge and disk colours suggests that the process which truncated star formation in the bulge did not destroy the disk.  

The above arguments provide support for bulge growing through recurrent minor merger episodes over long timescales, given that minor mergers do not destroy the original disks. The accretion of a satellite leads to bulge growth caused by the combined effects of satellite mass deposition and inward transport of disk material to the region of the bulge during the satellite decay. This process increases simultaneously the $B/D$ ratio and the bulge S\'ersic index $n$. Because the effects of merging are probably cumulative, bulge growth and $n$ increase would be progressive, leading to a continous evolution of disk galaxies toward earlier Hubble types \citep{aguerribulges01,Bournaud05,elichebulges06}. As a consequence of this systematic external driven secular evolution the final bulge-disk structure is strongly coupled.  Minor mergers also contribute to truncating star formation through redistribution of gas \citep{Rothberg06}. While the literature has more often emphasized the triggering of nuclear starbursts \citep{BarnesHernquist91,MihosHernquist96}, gas redistribution must occur throughout the galaxy and a progressive decrease in the disk star formation is a natural by-product.  

Recent studies point to the possibility that AGN feedback play an important role in the quenching of the star formation by regulating star formation in galaxies \citep{Kauffmann03agn,Nandra07,Georgakakis08}. Intense star formation, caused i.e. by minor merger, may trigger the AGN activity via starburst winds fueling the central black hole \citep{Younger08}. Once the AGN activity is powerful enough, it might suppress star formation via disruption of gas cooling or via driving gas out \citep{Hopkins05,Springel05}, in a way that star formation progressively ceases from inside out. This scenario could explain the small negative colour gradients in galaxies with prominent bulges.

\section{Conclusions}
\label{sec:conclusions}

\begin{enumerate}

\item $60\%$ of galaxies with high central brightness excess above the outer exponential ($\eta > 1$) concentrate strongly at 'very red' colours, $(U-B) > 0.25$, defining a red sequence, which fits well passive evolution models of different ages. The remainder $40\%$ span over a bluer and wider colour range.

\item Galaxies without central brightness excess ($\eta < 1$) show bluer colours and lack a red sequence. Galaxies show a colour bimodality that is related to the central brigthness excess, in the sense that more concentrated galaxies are redder. 

\item Nuclear and global colours are very similar to each other: redder bulges are embedded in redder galaxies. Disk colours also correlate with bulge colours. As a result, the colour difference between bulges and disks is smaller than that between bulges of different galaxies.  Colour profiles are smooth, and do not show a discontinuity at the bulge-disk boundary. 

\item The correlations between nuclear and disk regions are fulfilled for all types of galaxies, independently of the central brightness excess and the colour. The colour profiles are equally smooth in the samples with ($\eta > 1$) and without ($\eta > 1$) central concentration. 

\item A major merger origin for bulges, as proposed by others, needs to be able to account for the similarities in bulge and disk colours.  Such similarities lend support to models with coeval growth of bulges and disks, and to bulge formation models that do not destroy a pre-existing disk.   

\end{enumerate}

\begin{acknowledgements}
We thank the anonymous referee for suggestions that improved the paper and Mercedes Prieto, Peter Erwin, Ignacio Trujillo, Carmen Eliche-Moral, Carlos L\'opez, David Abreu, Marc Vallb\'e, David Crist\'obal, Enrique Garc\'\i a-Dab\'o, Alfonso Arag\'on-Salamanca, David Koo, and Reynier Peletier, for useful discussions. 
This work was supported by the Spanish Programa Nacional de Astronom\'\i a y Astrof\'\i sica through project number AYA2006-12955. 
Some of the data presented in this paper were obtained from the Multimission Archive at the Space Telescope Science Institute (MAST). STScI is operated by the Association of Universities for Research in Astronomy, Inc., under NASA contract NAS5-26555. Support for MAST for non-HST data is provided by the NASA Office of Space Science via grant NAG5-7584 and by other grants and contracts.
This work uses data obtained with support of the National Science Foundation grants AST 95-29028 and AST 00-71198.
\end{acknowledgements}

\bibliographystyle{aa}
\bibliography{9407refs}

\begin{thebibliography}{40}
\expandafter\ifx\csname natexlab\endcsname\relax\def\natexlab#1{#1}\fi

\bibitem[{{Aguerri} {et~al.}(2001){Aguerri}, {Balcells}, \&
  {Peletier}}]{aguerribulges01}
{Aguerri}, J.~A.~L., {Balcells}, M., \& {Peletier}, R.~F. 2001, \aap, 367, 428

\bibitem[{{Balcells} \& {Peletier}(1994)}]{Balcells94}
{Balcells}, M. \& {Peletier}, R.~F. 1994, AJ, 107, 135

\bibitem[{{Barnes} \& {Hernquist}(1991)}]{BarnesHernquist91}
{Barnes}, J.~E. \& {Hernquist}, L.~E. 1991, \apjl, 370, L65

\bibitem[{{Baugh} {et~al.}(1996){Baugh}, {Cole}, \& {Frenk}}]{Baugh96}
{Baugh}, C.~M., {Cole}, S., \& {Frenk}, C.~S. 1996, MNRAS, 283, 1361

\bibitem[{{Bolzonella} {et~al.}(2000){Bolzonella}, {Miralles}, \&
  {Pell{\'o}}}]{hyperz}
{Bolzonella}, M., {Miralles}, J.-M., \& {Pell{\'o}}, R. 2000, \aap, 363, 476

\bibitem[{{Bournaud} {et~al.}(2005){Bournaud}, {Jog}, \& {Combes}}]{Bournaud05}
{Bournaud}, F., {Jog}, C.~J., \& {Combes}, F. 2005, \aap, 437, 69

\bibitem[{{Carollo} {et~al.}(2001){Carollo}, {Stiavelli}, {de Zeeuw}, {Seigar},
  \& {Dejonghe}}]{Carollo01}
{Carollo}, C.~M., {Stiavelli}, M., {de Zeeuw}, P.~T., {Seigar}, M., \&
  {Dejonghe}, H. 2001, \apj, 546, 216

\bibitem[{{Chabrier}(2003)}]{Chabrier03}
{Chabrier}, G. 2003, \pasp, 115, 763

\bibitem[{{Cole} {et~al.}(2000){Cole}, {Lacey}, {Baugh}, \& {Frenk}}]{Cole00}
{Cole}, S., {Lacey}, C.~G., {Baugh}, C.~M., \& {Frenk}, C.~S. 2000, MNRAS, 319,
  168

\bibitem[{{Dominguez-Palmero} {et~al.}(2008){Dominguez-Palmero}, {Balcells},
  {Erwin}, {Prieto}, {Cristobal-Hornillos}, {Eliche-Moral}, \&
  R.}]{Dominguez08I}
{Dominguez-Palmero}, L., {Balcells}, M., {Erwin}, P., {et~al.} 2008, A\&A in
  press

\bibitem[{{Driver} {et~al.}(2006){Driver}, {Allen}, {Graham}, {Cameron},
  {Liske}, {Ellis}, {Cross}, {De Propris}, {Phillipps}, \& {Couch}}]{Driver06}
{Driver}, S.~P., {Allen}, P.~D., {Graham}, A.~W., {et~al.} 2006, \mnras, 368,
  414

\bibitem[{{Eisenhardt} {et~al.}(2007){Eisenhardt}, {De Propris}, {Gonzalez},
  {Stanford}, {Wang}, \& {Dickinson}}]{Eisenhardt07}
{Eisenhardt}, P.~R., {De Propris}, R., {Gonzalez}, A.~H., {et~al.} 2007, \apjs,
  169, 225

\bibitem[{{Eliche-Moral} {et~al.}(2006){Eliche-Moral}, {Balcells}, {Aguerri},
  \& {Gonz{\'a}lez-Garc{\'{\i}}a}}]{elichebulges06}
{Eliche-Moral}, M.~C., {Balcells}, M., {Aguerri}, J.~A.~L., \&
  {Gonz{\'a}lez-Garc{\'{\i}}a}, A.~C. 2006, \aap, 457, 91

\bibitem[{{Ellis} {et~al.}(2001){Ellis}, {Abraham}, \& {Dickinson}}]{Ellis01}
{Ellis}, R.~S., {Abraham}, R.~G., \& {Dickinson}, M. 2001, ApJ, 551, 111

\bibitem[{{Georgakakis} {et~al.}(2008){Georgakakis}, {Nandra}, {Yan},
  {Willner}, {Lotz}, {Pierce}, {Cooper}, {Laird}, {Koo}, {Barmby}, {Newman},
  {Primack}, \& {Coil}}]{Georgakakis08}
{Georgakakis}, A., {Nandra}, K., {Yan}, R., {et~al.} 2008, \mnras, 385, 2049

\bibitem[{{Hammer} {et~al.}(2001){Hammer}, {Gruel}, {Thuan}, {Flores}, \&
  {Infante}}]{Hammer01}
{Hammer}, F., {Gruel}, N., {Thuan}, T.~X., {Flores}, H., \& {Infante}, L. 2001,
  \apj, 550, 570

\bibitem[{{Hogg} {et~al.}(2004){Hogg}, {Blanton}, {Brinchmann}, {Eisenstein},
  {Schlegel}, {Gunn}, {McKay}, {Rix}, {Bahcall}, {Brinkmann}, \&
  {Meiksin}}]{Hogg04}
{Hogg}, D.~W., {Blanton}, M.~R., {Brinchmann}, J., {et~al.} 2004, \apjl, 601,
  L29

\bibitem[{{Hopkins} {et~al.}(2005){Hopkins}, {Hernquist}, {Cox}, {Di Matteo},
  {Martini}, {Robertson}, \& {Springel}}]{Hopkins05}
{Hopkins}, P.~F., {Hernquist}, L., {Cox}, T.~J., {et~al.} 2005, \apj, 630, 705

\bibitem[{{Kauffmann}(1996)}]{Kauffmann96bul}
{Kauffmann}, G. 1996, \mnras, 281, 487

\bibitem[{{Kauffmann} {et~al.}(2003){Kauffmann}, {Heckman}, {Tremonti},
  {Brinchmann}, {Charlot}, {White}, {Ridgway}, {Brinkmann}, {Fukugita}, {Hall},
  {Ivezi{\'c}}, {Richards}, \& {Schneider}}]{Kauffmann03agn}
{Kauffmann}, G., {Heckman}, T.~M., {Tremonti}, C., {et~al.} 2003, \mnras, 346,
  1055

\bibitem[{{Koo} {et~al.}(2005{\natexlab{a}}){Koo}, {Datta}, {Willmer},
  {Simard}, {Tran}, \& {Im}}]{Koo05HDF}
{Koo}, D.~C., {Datta}, S., {Willmer}, C.~N.~A., {et~al.} 2005{\natexlab{a}},
  \apjl, 634, L5

\bibitem[{{Koo} {et~al.}(2005{\natexlab{b}}){Koo}, {Simard}, {Willmer},
  {Gebhardt}, {Bouwens}, {Kauffmann}, {Crosby}, {Faber}, {Harker},
  {Sarajedini}, {Vogt}, {Weiner}, {Phillips}, {Im}, \& {Wu}}]{Koo05}
{Koo}, D.~C., {Simard}, L., {Willmer}, C.~N.~A., {et~al.} 2005{\natexlab{b}},
  \apjs, 157, 175

\bibitem[{{Kormendy} \& {Kennicutt}(2004)}]{Kormendy04}
{Kormendy}, J. \& {Kennicutt}, Jr., R.~C. 2004, \araa, 42, 603

\bibitem[{{MacArthur} {et~al.}(2008){MacArthur}, {Ellis}, {Treu}, {U}, {Bundy},
  \& {Moran}}]{MacArthur07}
{MacArthur}, L.~A., {Ellis}, R.~S., {Treu}, T., {et~al.} 2008, \apj, 680, 70

\bibitem[{{Marleau} \& {Simard}(1998)}]{Marleau98}
{Marleau}, F.~R. \& {Simard}, L. 1998, ApJ, 507, 585

\bibitem[{{Menanteau} {et~al.}(2001){Menanteau}, {Abraham}, \&
  {Ellis}}]{Menanteau01}
{Menanteau}, F., {Abraham}, R.~G., \& {Ellis}, R.~S. 2001, \mnras, 322, 1

\bibitem[{{Mihos} \& {Hernquist}(1996)}]{MihosHernquist96}
{Mihos}, J.~C. \& {Hernquist}, L. 1996, \apj, 464, 641

\bibitem[{{M{\"o}llenhoff} {et~al.}(2006){M{\"o}llenhoff}, {Popescu}, \&
  {Tuffs}}]{Moellenhoff06}
{M{\"o}llenhoff}, C., {Popescu}, C.~C., \& {Tuffs}, R.~J. 2006, \aap, 456, 941

\bibitem[{{Nandra} {et~al.}(2007){Nandra}, {Georgakakis}, {Willmer}, {Cooper},
  {Croton}, {Davis}, {Faber}, {Koo}, {Laird}, \& {Newman}}]{Nandra07}
{Nandra}, K., {Georgakakis}, A., {Willmer}, C.~N.~A., {et~al.} 2007, \apjl,
  660, L11

\bibitem[{{Peletier} \& {Balcells}(1996)}]{Peletier96}
{Peletier}, R.~F. \& {Balcells}, M. 1996, AJ, 111, 2238

\bibitem[{{Peletier} {et~al.}(1999){Peletier}, {Balcells}, {Davies},
  {Andredakis}, {Vazdekis}, {Burkert}, \& {Prada}}]{Peletier99}
{Peletier}, R.~F., {Balcells}, M., {Davies}, R.~L., {et~al.} 1999, MNRAS, 310,
  703

\bibitem[{{Rothberg} \& {Joseph}(2006)}]{Rothberg06}
{Rothberg}, B. \& {Joseph}, R.~D. 2006, \aj, 131, 185

\bibitem[{{Schulz} {et~al.}(2003){Schulz}, {Fritze-v.~Alvensleben}, \&
  {Fricke}}]{Schulz03}
{Schulz}, J., {Fritze-v.~Alvensleben}, U., \& {Fricke}, K.~J. 2003, \aap, 398,
  89

\bibitem[{{Scodeggio}(2001)}]{Scodeggio01}
{Scodeggio}, M. 2001, \aj, 121, 2413

\bibitem[{{Simard} {et~al.}(2002){Simard}, {Willmer}, {Vogt}, {Sarajedini},
  {Phillips}, {Weiner}, {Koo}, {Im}, {Illingworth}, \& {Faber}}]{Simard02}
{Simard}, L., {Willmer}, C.~N.~A., {Vogt}, N.~P., {et~al.} 2002, \apjs, 142, 1

\bibitem[{{Springel} {et~al.}(2005){Springel}, {Di Matteo}, \&
  {Hernquist}}]{Springel05}
{Springel}, V., {Di Matteo}, T., \& {Hernquist}, L. 2005, \mnras, 361, 776

\bibitem[{{Tuffs} {et~al.}(2004){Tuffs}, {Popescu}, {V{\"o}lk}, {Kylafis}, \&
  {Dopita}}]{Tuffs04}
{Tuffs}, R.~J., {Popescu}, C.~C., {V{\"o}lk}, H.~J., {Kylafis}, N.~D., \&
  {Dopita}, M.~A. 2004, \aap, 419, 821

\bibitem[{{Weinberg}(1972)}]{Weinberg72}
{Weinberg}, S. 1972, {Gravitation and Cosmology: Principles and Applications of
  the General Theory of Relativity} (pp.~688.~ISBN 0-471-92567-5.~Wiley-VCH,
  July 1972.)

\bibitem[{{Younger} {et~al.}(2008){Younger}, {Hopkins}, {Cox}, \&
  {Hernquist}}]{Younger08}
{Younger}, J.~D., {Hopkins}, P.~F., {Cox}, T.~J., \& {Hernquist}, L. 2008,
  ArXiv e-prints, 804

\bibitem[{{Zoccali} {et~al.}(2003){Zoccali}, {Renzini}, {Ortolani}, {Greggio},
  {Saviane}, {Cassisi}, {Rejkuba}, {Barbuy}, {Rich}, \& {Bica}}]{Zoccali03}
{Zoccali}, M., {Renzini}, A., {Ortolani}, S., {et~al.} 2003, \aap, 399, 931

\end{thebibliography}

\end{document}